# Biquaternionic Model of Electro-Gravimagnetic Field, Charges and Currents. Law of Inertia

L. A. Alexeyeva

Institute of Mathematics and Mathematical Modeling, Almaty, Kazakhstan
Email: alexeeva@math.kz

## Abstract

One the base of Maxwell and Dirac equations the one biquaternionic model of electro-gravimagnetic (EGM) fields is considered. The closed system of biquaternionic wave equations is constructed for determination of free system of electric and gravimagnetic charges and currents and generated by them EGM-field. By using generalized functions theory the fundamental and regular solutions of this system are determined and some of them are considered (spinors, plane waves, shock EGM-waves and others). The properties of these solutions are investigated.

## Keywords

## 1. Introduction

The one biquaternionic model of electro-gravimagnetic (EGM) fields and their interaction was elaborated by author in [1] [2]. There the fields analogues of three Newton laws for densities of mass and electric charge and current, acting forces and their powers have been built.

Here we consider the EGM-field created by free system of mass, charges and currents and their motion under action only internal electric and gravimagnetic tensions. In this model gravitational field (which is potential) is united with magnetic field (which is torsional) what gives possibility to enter gravimagnetic tension, charge and current. Lasts contain gravitational mass and their motion but not only them. Also here the new scalar $\alpha$-field of attraction-resistance is entered and their existence is justified. This phenomenon explains existence of longitudinal EM-wave which is observed in practice.

We use here differential algebra of biquaternions in hamiltonian form which more full were described in [3]. The scalar-vector form of biquaternion representation (*hamiltonian biform*) is very demonstrative and strangely adapted for writing the physical values and equations.

The base of this model is generalized biquaternionic form of Maxwell equations which includes differential part of Dirac operator [4]. From this form follow bigradiental representation of electric and gravimagnetic charges and currents. Differential operator *bigradient* is the generalization of gradient operator on the space of biquaternions which characterizes a direction of more extensive change of biquaternionic functions.

## 2. Biwave Equation and Its Solutions

To use biquaternions algebra we give some definitions. We enter on $\mathbb{M} = \{(\tau, x)\}$ (*Minkowski space*) the functional space of biquaternions in *hamiltonian form*:

$$\mathbb{B}(\mathbb{M}) = \{F = f(\tau, x) + F(\tau, x)\}$$

$f$ is a complex function, $F = \sum_{j=1}^{3} F_j e_j$ — a three-dimensional complex vector-function; $x = \sum_{j=1}^{3} x_j e_j$, $e_0 = 1, e_1, e_2, e_3$ are *basic elements* [3]. We assume $f(\tau, x), F_j(\tau, x)$ are locally integrable and differentiable on $\mathbb{M}$ or, in general case, they are generalized functions [5].

Summation and quaternionic multiplication are defined as

$$\alpha \mathbf{F} + \beta \mathbf{B} = \alpha(f + F) + \beta(b + B) = (\alpha f + b\beta) + (\alpha F + \beta B),$$

$$\mathbf{F} \circ \mathbf{B} = (f + F) \circ (b + B) = fb - (F, B) + fB + bF + [F, B],$$

where $(F, B) = F_j B_j, [F, B] = \epsilon_{jkl} e_j F_k B_l$ are usual scalar and vector productions in $R^3$ (here over repeated indexes there are summation from 1 to 3), $\epsilon_{jkl}$ is Levi-Civita symbol.

The norm and pseudonorm of Bq. are denoted

$$\|\mathbf{F}\| = \sqrt{|f|^2 + \|F\|^2}, \quad \langle\langle \mathbf{F} \rangle\rangle = \sqrt{|f|^2 - \|F\|^2}, \quad |f|^2 = f\bar{f}, \quad \|F\|^2 = (F, \bar{F}).$$

We'll use *convolution of biquaternions*:

$$\mathbf{F} * \mathbf{B} = f * b - (F_j * B_j) + f * B_j e_j + b * F_j e_j + \epsilon_{jkl} e_j F_k * B_l.$$

For regular components a convolution has the form:

$$f * b = \int_{\mathbb{M}} f(\tau - t, x - y) b(t, y) \mathrm{d}t \mathrm{d}y_1 \mathrm{d}y_2 \mathrm{d}y_3,$$

to take a convolution for singular generalized function and conditions of convolution existence see [5].

*Mutual bigradients* $\nabla^+, \nabla^-$ are the differential operators

$$\nabla^\pm \mathbf{B} = (\partial_\tau \pm i\nabla) \circ (b + B) = \partial_\tau b \mp i(\nabla, B) \pm i\nabla b \pm \partial_\tau B \pm i[\nabla, B]$$
$$= \partial_\tau b \mp i\,\mathrm{div}B \pm i\,\mathrm{grad}\,b \pm \partial_\tau B \pm i\,\mathrm{rot}B$$

Composition of mutual bigradients gives classic wave operator:

$$\nabla^+ \nabla^- \mathbf{B} = \nabla^- \nabla^+ \mathbf{B} = \partial_\tau^2 - \Delta = \square \quad (\text{dalambertian})$$

It gives possibility easy to construct the solutions of biquaternionic wave equation (*biwave* Equation)

$$\nabla^+ \mathbf{B} = \mathbf{G} \Rightarrow \square \mathbf{B} = \nabla^- \mathbf{G} \qquad (1)$$

which are presented in the form of the convolution:

$$\mathbf{B} = \nabla^-(\psi * \mathbf{G} + \mathbf{F^0})$$
$$= \nabla^- \left( \int_{\|y-x\| \le \tau} \frac{\mathbf{G}(y, \tau - \|y-x\|)}{4\pi \|y-x\|} \mathrm{d}y_1 \mathrm{d}y_2 \mathrm{d}y_3 + \mathbf{F^0} \right) \qquad (2)$$

where

$$\psi = \frac{1}{4\pi\|x\|}\delta(\tau - \|x\|)$$

is the fundamental solution of D'Alember equation (*a simple layer on light cone*):

$$\Box\psi = \delta(\tau)\delta(x),$$

$\mathbf{F^0} = \psi_j^0 e_j$ is arbitrary solution of homogeneous D'Alember equation:

$$\Box\psi_j^0 = 0.$$

In formulae (2) the second equality is written for regular **G**.

We name

$\psi, \psi_j^0$ —*scalar potentials* of **B**,

$\Psi^0 = \psi_j^0 e_j$ —*vector potentials*

$\Psi^0 = \nabla^-(\psi_0^0 + \Psi^0)$ —*spinor*.

$\overline{\mathbf{F}} = \overline{f} + \overline{F}$ —*complex conjugate* to F;

$\mathbf{F}^* = \overline{f} - \overline{F}$ —*conjugate* to F.

If $\mathbf{F}^* = \mathbf{F}$, it is *selfconjugated* Bq. Selfconjugated Bq. has the form $\mathbf{F} = f + iF$, $f$ and $F$ have real values. For example

$$\mathbf{B} \circ \mathbf{B}^* = \|B\|^2 + 2iIm(\overline{b}B).$$

## 3. Characteristics of Electro-Gravimagnetic Field

Let introduce known and new physical values which characterize EGM-field, charges and currents:
- real vectors $E$ and $H$ are the tensions of electric and *gravimagnetic fields*;
- real scalars $\rho^E, \rho^H$ are the densities of electric and *gravimagnetic charges*;
- real vectors $j^E, j^H$ are the densities of electric and *gravimagnetic current*.

Here we united the gravitational field (which is *potential*) with magnetic field (which is *torsional*) in one *gravimagnetic* field $H$. Also we united mass current with magnetic currents. As well known classic electrodynamics refuse the existence of magnetic charges and currents. But here we'll show that $\rho^H$ and $j^H$ can the rights on the existence.

By using these values we introduce the complex characteristics of EGM-field:
- complex vector of *EGM-intensity*

$$A(\tau, x) = A^E + iA^H = \sqrt{\varepsilon}E + i\sqrt{\mu}H;$$

- complex charges field:

$$\rho(\tau, x) = -\rho^E/\sqrt{\varepsilon} + i\rho^H/\sqrt{\mu};$$

- complex currents field:

$$J(\tau, x) = -\sqrt{\mu}j^E + i\sqrt{\varepsilon}j^H;$$

- complex scalar *field of attraction-resistance*

$$\alpha(\tau, x) = \frac{ia_1}{\sqrt{\varepsilon}} + \frac{a_2}{\sqrt{\mu}}.$$

Here values $\varepsilon, \mu$ are constants of electric conductivity and magnetic permeability of corresponding EM-medium.

## 4. Biquaternions of Electro-Gravimagnetic Field

We construct the next Bqs. of EGM-field and charge-currents field (*CC-field*):

- *EGM-potential*

$$\mathbf{\Phi} = \phi + i\Psi,$$

- *EGM-intensity*

$$\mathbf{A}(\tau, x) = i\alpha(\tau, x) + A(\tau, x),$$

- *charge-current*

$$\Theta(\tau, x) = i\rho(\tau, x) + J(\tau, x),$$

- *energy-pulse* of EGM-field

$$\Xi(\tau, x) = 0.5\mathbf{A}^* \circ \mathbf{A} = W(\tau, x) + iP(\tau, x).$$

In case $\alpha = 0$ here you see the energy density $W$ and Pointing vector $P$ of EM-field:

$$W = 0.5\left(\varepsilon \|E\|^2 + \mu\|H\|^2\right), \quad P = c^{-1}[E, H], \quad c = 1/\sqrt{\varepsilon\mu}, \tag{3}$$

$c$ is light speed.

By analogue we enter biquaternion of *energy-pulse* of charge-current field (*CC-field*)

$$\Omega(\tau, x) = 0.5\Theta^* \circ \Theta = \varpi(\tau, x) + i\Pi(\tau, x).$$

If to calculate

$$\Omega(\tau, x) = \left(\frac{|\rho_E|}{\varepsilon} + \frac{|\rho_H|}{\mu} + Q\right) + i\left(P_J - \sqrt{\frac{\mu}{\varepsilon}}\rho^E j^E - \sqrt{\frac{\varepsilon}{\mu}}\rho^H j^H\right) \tag{4}$$

$\varpi$ contains $|\rho|$ and energy density of currents:

$$Q = 0.5\|J\|^2 = 0.5\left(\mu\|j^E\|^2 + \varepsilon\|j^H\|^2\right)$$

where the first summand includes Joule heat of electric current; second one includes energy density of gravimagnetic current, which contains kinetic energy of mass current. Here vector $P_J$ is analogue of Pointing vector, but for the current:

$$P_J = 0.5iJ \times \bar{J} = c^{-1}\left[j^H, j^E\right]$$

$P_J = 0$ only if gravimagnetic and electrical currents are parallel or one from them is equal 0.

## 5. Maxwell-Dirac Equation of EGM-Field

### 5.1. Connection between EGM-Field, Charges and Currents

*Postulate* 1. Connection between EGM-intensity and charge-current is bigradiental:

$$\nabla^+ \mathbf{A} = \Theta(\tau, \mathbf{x}) \tag{5}$$

This assumption follows from Maxwell equations.

In particulary by $\alpha = 0$ from here follow the known Hamiltonian form of Maxwell equations [6]:

$$divA + \rho(\tau, x) = 0, \quad \partial_\tau A + irotA + J(\tau, x) = 0. \tag{6}$$

As

$$\Box \mathbf{A} = \nabla^- \Theta$$

we have from here the known formulas for electric charge and currents:

$$\partial_\tau \rho + div J = 0, \quad \Box A = \nabla\rho + \partial_\tau J - i rot J. \tag{7}$$

From (5) for real and imaginary part we get
**generalized Maxwell equations**:

$$rotH - \varepsilon \frac{\partial E}{\partial t} + c^2 grad\, \alpha_1 = j^E(\tau, x),$$
$$rotE + \mu \frac{\partial H}{\partial t} - c^2 grad\, \alpha_2 = j^H(\tau, x),$$
$$\varepsilon divE + \partial_t \alpha_1 = \rho^E(\tau, x),$$
$$-\mu divH + \partial_t \alpha_2 = \rho^H(\tau, x)$$

(8)

which, when $\alpha = 0, \rho^H = 0, j^H = 0$, coincides with
*classic Maxwell equations*:

$$rotH - \varepsilon \frac{\partial E}{\partial t} = j^E(\tau, x),$$
$$rotE + \mu \frac{\partial H}{\partial t} = 0,$$
$$\varepsilon divE = \rho^E(\tau, x),$$
$$-\mu divH = 0.$$

(9)

When EGM-field and charge-currents are independent on time, we get from ((8)
*equations for stationary charges and currents*:

$$rotH + c^2 grad\, \alpha_1 = j^E(, x),$$
$$rotE - c^2 grad\, \alpha_2 = j^H(x),$$
$$\varepsilon divE = \rho^E(x),$$
$$\mu divH = -\rho^H(x).$$

(10)

From last two scalar equation easy to get the known Coulomb's equation for potential of *electrostatic field*. The second one gives the Poisson equation for potential of *Newton gravitational fields* if to put $\rho^H(x)$ is the *mass density* and $\mu = \frac{1}{4\pi\gamma}$, where $\gamma$ is gravitational constant.

*Remark.* We must note that the first scalar equation of classic Maxwell Equation (9) (where electric charges can depend on time) contradict to wave nature of EM-field. But Equation (9) is true only if charge and currents are independent on time. The same one relates to Eq. for gravitational field which is true only for static mass.

All this confirm postulate 1, which shows, that
*charges and currents of EGM-field are physical appearance of bigradient of EGM-intensity*!

From here follow,
*if bigradient of EGM-intensity is equal to zero then charges and currents are absent*!

Equation (5) is generalization of Maxwell equation in biquaternions algebra. The differential operator corresponding to it coincides with the differential part of matrix operator of Dirac [4]. By this course Equation (5) we name *Maxwell-Dirac equation of EGM-field* or simply the *EGM-equation*.

EGM-equation is hyperbolic, and corresponding to it system of differential Equation (8) is hyperbolic and connected. It's known that classic system of Maxwell Equation (9) doesn't possess such properties.

### 5.2. Generalized Solutions of EGM-Equation

As Equation (5) is biwave equation, to construct its solution it's need to use formulae (2):

$$\mathbf{A} = \nabla^- (\psi * \Theta + \mathbf{A^0}) = \nabla^- \left( \int_{\|y-x\|\leq \tau} \frac{\Theta(y, \tau - \|y-x\|)}{4\pi\|y-x\|} dy_1 dy_2 dy_3 + \mathbf{A^0} \right)$$

(11)

According to (2) the scalar and vector parts of EGM-intensity have the form:

$$\alpha = -i\partial_\tau (\psi * \rho) + i\sum_{j=1}^{3} \partial_j \psi^j + \partial_\tau \psi^0$$

$$A = i\,grad\,(\psi * \rho) - \partial_\tau (\psi * J) + i\,rot\,(\psi * J) + i\sum_{j=1}^{3}\partial_\tau \psi^j e_j + rot\sum_{j=1}^{3}\psi^j e_j.$$

For classic Maxwell equation $\alpha = 0$. Hence, for this case,

$$i\partial_\tau (\psi * \rho) = i\sum_{j=1}^{3}\partial_j \psi^j + \partial_\tau \psi^0.$$

In particulary it's performed if $\rho = 0$ and there is *Lorentz calibrations* for potentials:

$$i\sum_{j=1}^{3}\partial_j \psi^j + \partial_\tau \psi^0 = 0.$$

But if $\rho \neq 0$ and there is Lorentz calibration then $\alpha \neq 0$ and scalar $\alpha$-field exists!

## 5.3. Shock EGM-Waves. Conditions on Wave Front

As EGM-equation (5) is hyperbolic, it has characteristic surface ($F$). Its equation has the form [3] [6]:

$$v_t^2 \left( v_t^2/c^2 - v_1^2 - v_2^2 - v_3^2 \right)^2 = 0,$$

where $(v, v_t) \triangleq (v_1, v_2, v_3, v_t)$ is a normal to $F$ in Minkowski space $\mathbb{M}$. On $F$ solutions of (5) can have *jump* of EGM-field $[\mathbf{A}]_F$. Such waves are named *shock EGM-waves*. Vector $v$ is a normal to a wave front $F_t$ in $R^3$, which is moving with light speed c. *Unit wave vector* $m \triangleq v/\|v\|$ is directed towards the front movement and satisfied to equation:

$$c = -\frac{v_t}{\|v\|}$$

In the space of distributions the classical solution of (5) (considered as generalized biquaternions $\hat{\mathbf{A}}$) satisfies to the following equation:

$$\nabla^+ \hat{\mathbf{A}} = \hat{\mathbf{\Theta}} + \left\{ v_\tau [\alpha]_F + ([A]_F, v) - iv_\tau [A]_F + v \times [A]_F + i[\alpha]_F v \right\} \delta_F$$

Here singular generalized function $[A]_F \delta_F$ is simple layer on $F$, gap $[A]_F$ is density of this layer. From here follow that $\hat{\mathbf{A}}$ is the generalized solution of (5) only if the next conditions on EGM-waves fronts are performed:

*conditions on EGM-wave front*:

$$\left( [A]_{F_t}, m \right) = [\alpha]_{F_t}, \quad [A]_{F_t} = im \times [A]_{F_t} - [\alpha]_{F_t} m. \tag{12}$$

From here follow the next conditions for real and imaginary parts.
On fronts of shock EGM-waves the gaps of intensities satisfy to the next conditions:

$$\left( [E]_{F_t}, m \right) = c[a_2], \quad \left( [H]_{F_t}, m \right) = c[a_1].$$

$$[E]_{F_t} = \sqrt{\frac{\mu}{\varepsilon}} [H]_{F_t} \times m - c[a_2]_{F_t} m,$$

$$[H]_{F_t} = -\sqrt{\frac{\varepsilon}{\mu}} [E]_{F_t} \times m - c[a_2]_{F_t} m.$$

Here the sigh "×" notes vector production.

You see that for generalized Maxwell-Dirac equations shock waves are not transversal. Only if $[\alpha]_{F_t} = 0$ then they are *transverse*:

$$\left( [E]_{F_t}, m \right) = 0, \left( [H]_{F_t}, m \right) = 0.$$

It's well known for EM-waves as generalized solutions of classic Maxwell Equation (9).

You see that the longitudinal components at the front of EGM-wave are connected with a jump of scalar $\alpha$-field, which describes property of attraction-resistance of EGM-field to the movement of *external* charges and currents. It was shown in [1] [2].

### 5.4. Spinors of EGM-Field

At absence of charges and currants EGM-field satisfies to homogeneous biwave equation

$$\nabla^+ \mathbf{A} = 0,$$

which solutions can be constructed by use spinors in (2).

We consider here some unconventional spinor which can explain longitudinal electromagnetic waves, which are observed in practice [7] [8].

*Plane spinors.* Let construct some plane waves generated by scalar potentials:

$$\psi_j = f(\eta), \quad \eta = (k,x) - \omega t, \quad \|k\| = \omega$$

where $f(\ldots)$ is arbitrary function which describes the plane wave, moving in direction of wave vector $k$ with speed $c = \omega/\|k\| = 1$.

1) *Longitudinal magnetic wave* in direction $H$:

$$\mathbf{A}^0 = \nabla^- \psi^0 = (\partial_\tau - i\nabla)(f(\eta) + 0) = -\omega f'(\eta) - ikf'(\eta)$$

2) *Longitudinal electric wave* in direction $E$:

$$\mathbf{A}^0 = \nabla^- \psi^0 = (\partial_\tau - i\nabla)(if(\eta) + 0) = -\omega f'(\eta) - kf'(\eta)$$

3) *Tesla's wave*—EM-wave in direction with torsion component $H$:

$$\mathbf{A}^1 = (\partial_\tau - i\nabla)(0 + f(\eta)e_1) = ik_1 f'(\eta) - \omega e_1 f'(\eta) - ik_3 e_2 f'(\eta) + ik_2 e_3 f'(\eta)$$

as $(E,H) = 0$ and

$$(E,k) = -\frac{\omega k_1}{\sqrt{\varepsilon}} f'(\eta), \quad (H,k) = \left(-\frac{k_3 k_2}{\sqrt{\mu}} + \frac{k_3 k_2}{\sqrt{\mu}}\right) f'(\eta) = 0$$

4) *Torsion wave*—EM-wave in direction $H$ with torsion component $E$.

$$\mathbf{A}^1 = \nabla^- \psi^1 e_1 = (\partial_\tau - i\nabla)(0 + if(\eta)e_1) = i\omega e_1 f'(\eta) + k_3 e_2 f'(\eta) - k_2 e_3 f'(\eta)$$

$$(H,k) = \frac{\omega k_1}{\sqrt{\mu}} f'(\eta), \quad (E,k) = 0, \quad (E,H) = 0$$

Here waves names correspond to ones in paper of V.A. Etkin [8].

## 6. Law of Inertia

### 6.1. Free Field. Analogue of First Newton Law

EGM-equation (5) gives possibility to construct EGM-intensity if charge-current are known. And vice versa if EGM-field is known its bigradient determine charges-currents.

Hence this equation and corresponding system are unclosed. To close this equation let assume that the next proposition in true.

Postulate 2. *If the charge-current are free* (*there are absence the action of external EGM-fields*) *then*

$$\nabla^- \Theta = \mathbf{0} \tag{13}$$

*which is equivalent to equations*:

$$\partial_\tau \rho + \operatorname{div} J = 0, \quad \partial_\tau J - i\operatorname{rot} J + \operatorname{grad} \rho = 0 \quad \Leftrightarrow$$

*or for real and complex parts*

$$\partial_t \rho^E + \text{div } j^E = 0, \quad \partial_\tau j^E = \sqrt{\varepsilon/\mu} \, \text{rot } j^H - c \, \text{grad } \rho^E,$$
$$\partial_t \rho^H + \text{div } j^H = 0, \quad \partial_\tau j^H = -\sqrt{\mu/\varepsilon} \, \text{rot } j^E - c \, \text{grad } \rho^H. \quad (14)$$

Here the scalar equations are the known *conservation laws* of the electric and gravimagnetic charges which must be executed at absence of external actions (influence).

The Equation (13) is field's analogue of the first Newton law (*inertia law*). The systems (14) show that gradients of charges and rotors of currents stipulate their motion and changing.

## 6.2. Generalized Solution of Free Charge-Current Equation

Free $\Theta$ is a spinor, an arbitrary solution of homogeneous biwave equation:

$$\Theta = \nabla^- \psi^0 + i \sum_{j=1}^{3} \nabla^- (\psi^j e_j), \quad (15)$$

$$\Box \psi^j = 0, \quad j = 0, 1, 2, 3, 4$$

where

$$\psi^j = \int_{R^3} \varphi^j(\xi) \exp(-i(\xi, x) - i\|\xi\|t) d\xi_1 d\xi_2 d\xi_3, \quad \forall \varphi^j(\xi) \in L_1(R^3) \quad (16)$$

Solutions of wave equation have been studied very well. We give some examples of solutions of this equation which may be used for construction of elementary particles and more complex matter.

1) *Harmonic spherical $\omega$-spinor*. Their potentials are presented in the form:

$$\psi^j(n, k, l) = j_n(r) P_k^l(\cos\theta) e^{i(l\varphi - \omega\tau)}, \quad (17)$$

$(r, \theta, \varphi)$ are spherical coordinates, $j_n$ are spherical Bessel functions, $P_k^l$ are associated Legendre polynomials.

2) *Spinors field*

$$\Theta = \nabla^- \left\{ \psi^0 + i \sum_{j=1}^{3} \psi^j e_j \right\} * \mathbf{C}(\tau, x). \quad (18)$$

3) *Fibers*

$$\Theta = \nabla^- \left\{ \psi^0 + i \sum_{j=1}^{3} \psi^j e_j \right\} * \mathbf{C}(\tau, x) \delta_l(x, \tau). \quad (19)$$

4) *Tissue*

$$\Theta = \nabla^- \left\{ \psi^0 + i \sum_{j=1}^{3} \psi^j e_j \right\} * \mathbf{C}(\tau, x) \delta_S(x, \tau). \quad (20)$$

5) *Body*

$$\Theta = \nabla^- \left\{ \psi^0 + i \sum_{j=1}^{3} \psi^j e_j \right\} * \mathbf{C}(\tau, x) H_S^-(x). \quad (21)$$

Here $\mathbf{C}(\tau, x)$ are arbitrary Bqs, which admit such convolutions. Generalized functions $\delta_l(x, \tau)$, $\delta_S(x, \tau)$ are simple layers on lines (*l*) and surfaces (*S*), $H_S^-(x)$ is characteristic function of domain which is bounded by *S*.

## 7. Closed System of Equations for EGM-Field, Charges and Currents

Equations ((5) and (13)) give full and closed system of hyperbolic type for determination field of free charge and currents and generated EGM-field, which we formulate as postulate 3.

Postulate 3. *The EGM-field of free charge-current are described by the next biquaternionic system*:

$$\nabla^+ \mathbf{A} = \Theta(\tau, \mathbf{x}),$$
$$\nabla^- \Theta = 0. \tag{22}$$

*Its hamiltonian form*

$$\partial_\tau \alpha - \operatorname{div} A = \rho,$$
$$\partial_\tau A + i \operatorname{rot} A - \operatorname{grad} \alpha = J(\tau, x), \tag{23}$$
$$\partial_\tau \rho + \operatorname{div} J = 0,$$
$$\operatorname{grad} \rho + \partial_\tau J - i \operatorname{rot} J = 0.$$

*Its differential equations system*

$$-\varepsilon \frac{\partial E}{\partial t} + \operatorname{rot} H + c^2 \operatorname{grad} \alpha_1 = j^E(\tau, x),$$
$$\mu \frac{\partial H}{\partial t} + \operatorname{rot} E - c^2 \operatorname{grad} \alpha_2 = j^H(\tau, x),$$
$$\partial_\tau j^E - \sqrt{\varepsilon/\mu} \operatorname{rot} j^H + c \operatorname{grad} \rho^E = 0$$
$$\partial_\tau j^H + \sqrt{\mu/\varepsilon} \operatorname{rot} j^E + c \operatorname{grad} \rho^H = 0. \tag{24}$$
$$\varepsilon \operatorname{div} E + \partial_t \alpha_1 = \rho^E(\tau, x),$$
$$-\mu \operatorname{div} H + \partial_t \alpha_2 = \rho^H(\tau, x)$$
$$\partial_t \rho^E + \operatorname{div} j^E = 0,$$
$$\partial_t \rho^H + \operatorname{div} j^H = 0.$$

Their general solutions are presented in the form (15), (11).

## 8. Conclusions

Here we consider the fields analogue of first Newton law which has been postulated not for material point but for distributed electric and gravimagnetic charges and currents. We show, that charges and currents of EGM-field are physical appearance of bigradient of EGM-intensity. If bigradient of EGM-intensity is equal to zero then charges and currents are absent.

From this model follow that charges $\rho^E$ and $\rho^H$ can be positive and negative. By stationary vibration (as follow from solution (17)) $\rho^E$ can transfer to $\rho^H$ and vice-versa. In static case gravimagnetic charge is like to mass. We suppose, a *density of mass* is $|\rho|$ which enters into the Equation (4) for energy density $\varpi$.

We introduced postulates for EGM-field on the base of generalization of biquaternionic form of Maxwell equations and obtained closed hyperbolic system which connects EGM-field, charges and currents in united system of equations. For this we enter new scalar $\alpha$-field of attraction-resistance which gives possibility to explain some physical phenomena which are observed in practice.

In particular, the solutions of EGM-field describe electric and gravimagnetic waves which, in general case, are not transversal and have longitudinal component. Longitudinal EM-waves are observed in practice but classic electrodynamics doesn't explain their existence.

Many interesting physical properties of this model appear by interaction of different system of charges and currents and their EGM-fields. Some of them were described in papers [1] [2].